\documentclass[jkps,twocolumn,twoside,fleqn,showpacs,showkeys]{revtex4}
\usepackage{graphicx}
\usepackage{amssymb}
\usepackage{amsmath}
\usepackage{bm}
\usepackage{picture}

\begin{document}
\setcounter{page}{0}
\title[]{Direct Observation and Analysis of Spin Dependent Transport
of Single Atoms in a 1D Optical Lattice}
\author{Micha{\l} \surname{Karski}}
\author{Leonid \surname{F\"orster}}
\author{Andrea \surname{Alberti}}
\author{Wolfgang \surname{Alt}}
\author{Artur \surname{Widera}}
\author{Dieter \surname{Meschede}}
\email{meschede@uni-bonn.de}
\thanks{Fax: +49-228-733474}
\affiliation{Institut f\"ur Angewandte Physik, Universit\"at Bonn}
\author{Jai-Min \surname{Choi}}
\affiliation{Department of Science Education, Chonbuk National
University, Jeonju 561-756, Korea}


\begin{abstract}
We have directly observed spin-dependent transport of single
cesium atoms in a 1D optical lattice. A superposition of two
circularly polarized standing waves is generated from two counter
propagating, linearly polarized laser beams. Rotation of one of
the polarizations by $\pi$ causes displacement of the $\sigma^{+}$
and the $\sigma^{-}$-lattices by one lattice site. Unidirectional
transport over several lattice sites is achieved by rotating the
polarization back and forth and flipping the spin after each
transport step. We have analyzed the transport efficiency over 10
and more lattice sites, and we discuss and quantify the relevant
error sources.
\end{abstract}



\maketitle
\section{Introduction}
Controlled transport of atoms stored in optical lattices is a central process in the quest for coherent atom-atom interactions. Such interactions are at the heart of creating and manipulating quantum multi-particle systems for, \textit{e.g.}, quantum simulations. Spin-dependent transport --- where the direction of transport depends on the spin state --- opens especially interesting routes because it allows the creation of spin-position entangled quantum states as precursors for correlated many-particle states. A concept for such a transport with neutral atoms stored in an optical lattice was proposed by Deutsch and Jessen and by Jaksch \textit{et~al.}\cite{Deu98, Jak99} and first demonstrated by Mandel \textit{et~al.}\cite{Man03} with a sample of ultracold rubidium atoms in a Mott insulator state, where the signature of transport was observed in momentum space.

Here, we report a direct observation of spin-dependent transport in a 1D optical lattice through fluorescence imaging. In contrast to the ``top-down'' approach of Ref.~\citealp{Man03} we use single cesium atoms prepared in low-energy thermal states in deep lattice sites\cite{Alt03}.

\section{Spin-dependent transport}
\begin{figure}
    \centering
\includegraphics[width=0.8\columnwidth]{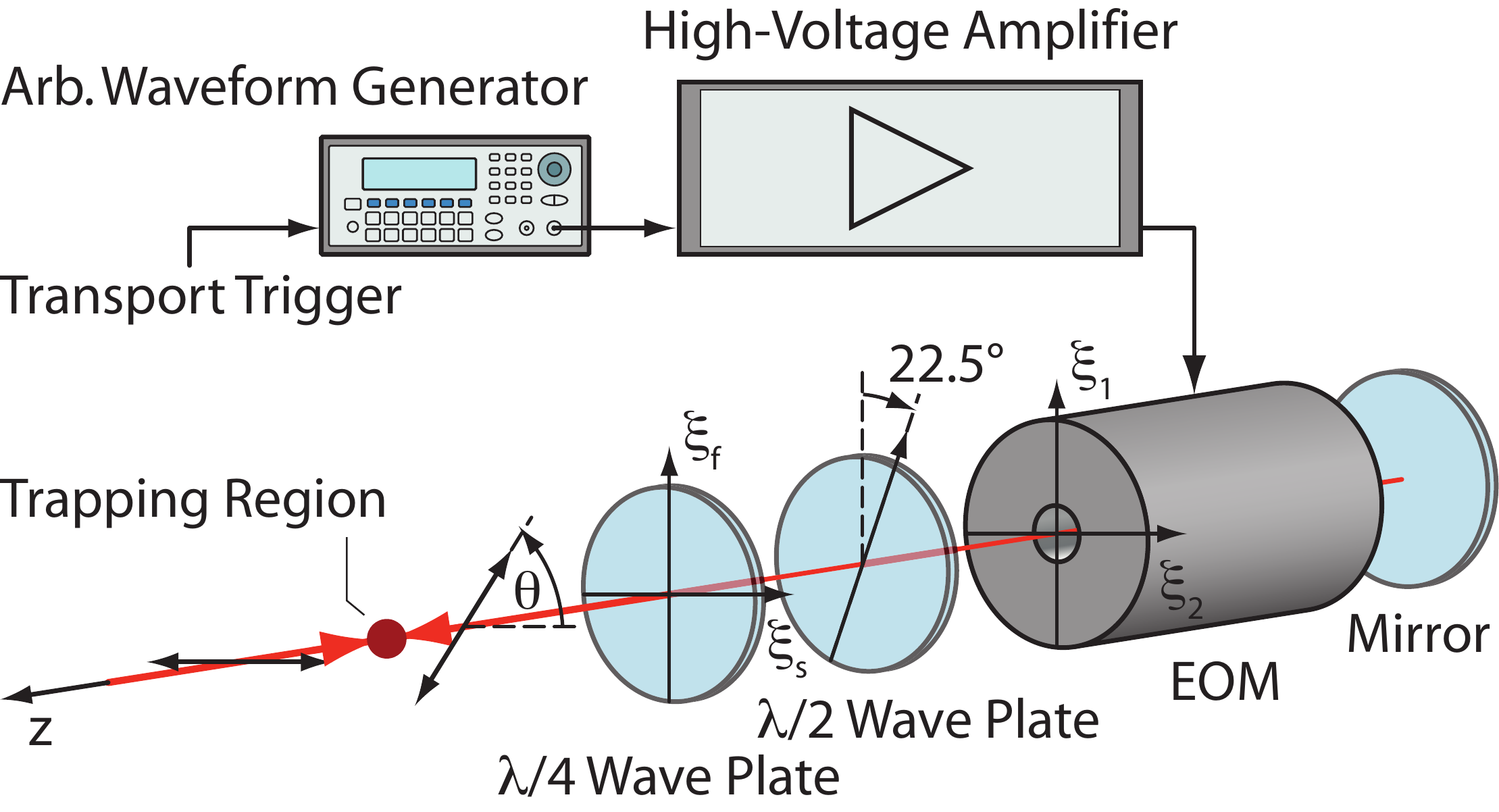}
\caption{(Color online) Experimental setup to control the rotation angle $\theta$ between the polarization vectors of an incident and a retro-reflected linearly polarized beam using an electro-optical modulator (EOM). Both principle axes of the quarter-wave plate, $\xi_{\text{f}}$ and $\xi_{\text{s}}$ (fast and slow axes), and those of the EOM, $\xi_{1}$ and $\xi_{2}$, are parallel and perpendicular to the polarization vector of the incident beam, respectively. A~half-wave plate in between is used to rotate the incoming polarization to $45^{\circ}$ with respect to the principal axes of the EOM.}\label{fig1}
\end{figure}

In a spin-dependent transport, the shift direction of a trapped atom along the lattice axis is determined by its internal spin (or qubit) state $\left|s\right>=\left|\uparrow\right>$ or $\left|\downarrow\right>$. Such transport can be realized using two counterpropagating linearly polarized laser beams in a lin-$\theta$-lin configuration\cite{Jak99}, in which the rotation angle $\theta$ of the polarization vector of one of the laser beams is continuously varied (see Fig.~\ref{fig1}). The resulting light field can be decomposed into a $\sigma^{+}$ and a $\sigma^{-}$ circularly polarized standing wave, contributing to the trapping potential by $U_{+}(z,\theta)=U_{0}\cos(k z-\theta/2)$ and $U_{-}(z,\theta)=U_{0}\cos(k z+\theta/2)$, respectively. Here, $k=2\pi/\lambda$ is the wave vector component along the lattice axis, and $U_{0}$ is the depth of the potential wells. By varying the rotation angle $\theta$, both standing waves and their contributions to the trapping potential are spatially shifted in opposite directions by a distance of $z_{\pm}=\pm\theta/\pi\cdot\lambda/4$, resulting in a relative displacement of $\Delta z=z_{+}-z_{-}=\theta/\pi\cdot\lambda/2$ and an overlapping at $\theta/\pi=0,\pm 1,\pm 2,\ldots$. For the outermost Zeeman sublevels of the cesium hyperfine ground state manifold, a magic wavelength $\lambda_{\text{m}}$ in between the $\text{D}_{1}$ and the $\text{D}_{2}$ lines can be found such that one spin state only experiences the $U_{+}(z,\theta)$ component of the trapping potential while the other spin state is mainly affected by the $U_{-}(z,\theta)$ component.

In our experiment, the rotation angle $\theta$ is
voltage-controlled using an electro-optical modulator (EOM) (see
Fig.~\ref{fig1}). The EOM is driven by a high-voltage amplifier
with a $-3\,\text{dB}$ bandwidth of $370\,\text{kHz}$ and a
maximum output voltage of $750\,\text{V}$, limiting the rotation
angle range to $0\leq\theta\leq 1.7\pi$. For the spin (or qubit)
states, we use
$\left|\uparrow\right>=\left|F=4,m_{\text{F}}=4\right>$ and
$\left|\downarrow\right>=\left|F=3,m_{\text{F}}=3\right>$. For the
optical lattice, we choose
$\lambda\equiv\lambda_{\text{m}}=865.9\,\text{nm}$ and
$U_{0}/k_{\text{B}}=80\,\mu\text{K}$, resulting in spin-dependent
trapping potentials $U_{\uparrow}(z,\theta)=U_{+}(z,\theta)$ and
$U_{\downarrow}(z,\theta)=\tfrac{1}{8}U_{+}(z,\theta)+
\tfrac{7}{8}U_{-}(z,\theta)$ with axial and radial trapping
frequencies of $\omega_{\text{ax}}=2\pi\times 115\,\text{kHz}$ and
$\omega_{\text{rad}}=2\pi\times 1\,\text{kHz}$, respectively. Due
to the residual contribution of the $U_{+}(z,\theta)$ component to
the $U_{\downarrow}(z,\theta)$ potential, its shape changes in
depth and contrast during the shift, affecting the axial and the
radial trapping frequencies of atoms transported in state
$\left|\downarrow\right>$. Furthermore, the spatial shift of
$U_{\downarrow}(z,\theta)$ depends nonlinearly on the rotation
angle $\theta$, see, \textit{e.g.}, Ref.~\citealp{Deu98}.

\begin{figure}
    \centering
\includegraphics[width=0.8\columnwidth]{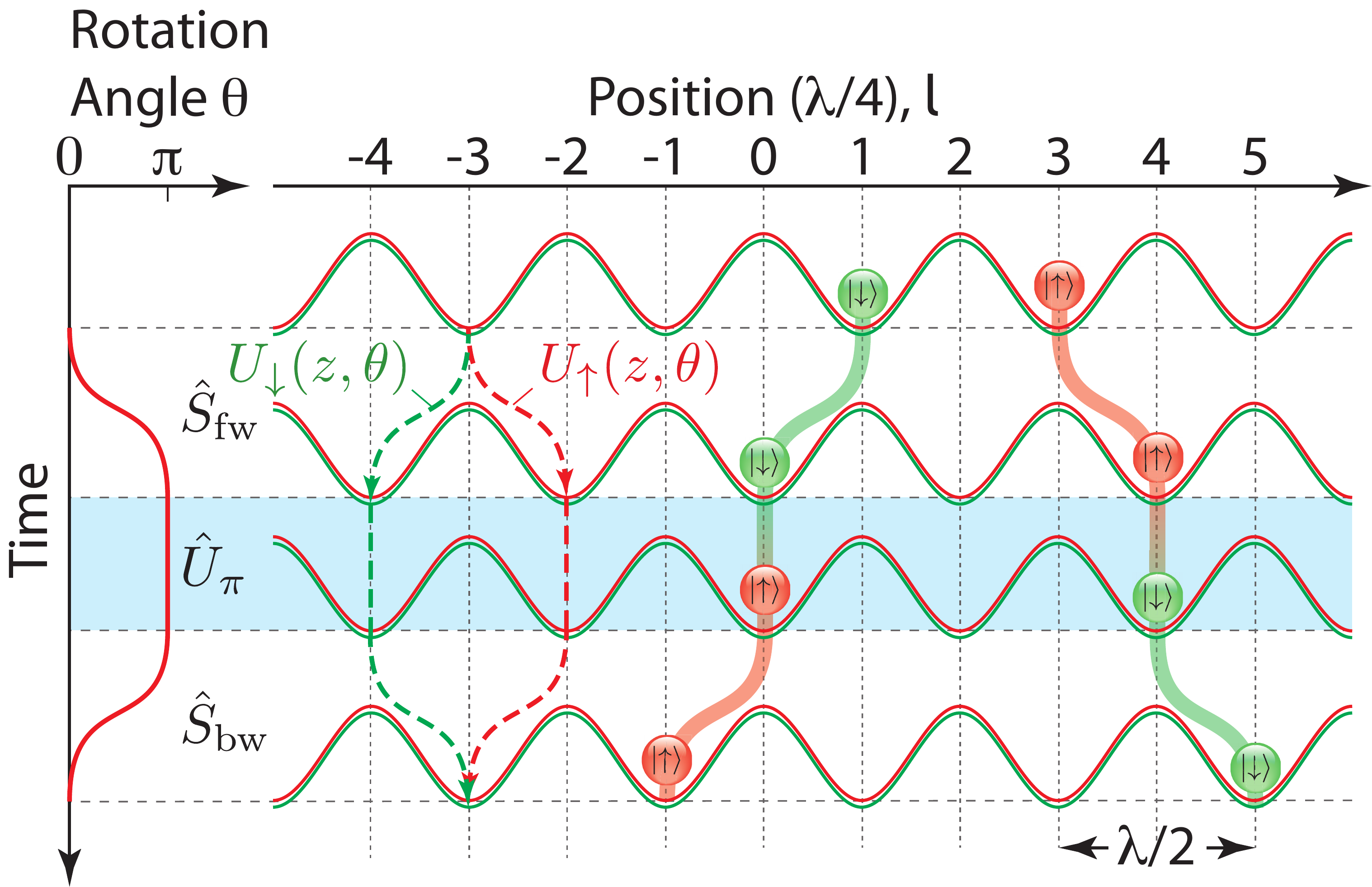}
\caption{(Color online) Schematic representation of unidirectional spin-dependent transport of atoms over distances of one lattice site. Shifts of the spin-dependent potentials $U_{\uparrow}(z,\theta)$ and $U_{\downarrow}(z,\theta)$ are indicated by dashed lines; and transport of atoms by shaded bold lines.}\label{fig2}
\end{figure}

The limited range of the rotation angle allows only two perfectly
overlapping trapping potential configurations: namely, $\theta=0$
and $\theta=\pi$, where
$U_{\uparrow}(z,\theta)=U_{\downarrow}(z,\theta)$. We, therefore,
define a single transport step as a shift by a distance of
$\pm\lambda/4$, \textit{i.e.}, from one overlapping configuration to
another. Taking into account the periodicity of the optical
lattice, it is convenient to discretize the position space along
the lattice axis in units of $\lambda/4$ (see Fig.~\ref{fig2}).
The spin-dependent shift operators then read
\begin{equation}\label{eq:shift-up-operator}
\hat{S}_{\text{fw}}:\begin{cases}\begin{array}{rcl}\left|\uparrow,l\right> &\to& e^{i\varphi_{\uparrow}}\left|\uparrow,l+1\right>\\
\left|\downarrow,l\right> &\to&e^{i\varphi_{\downarrow}}\left|\downarrow,l-1\right>\end{array}\end{cases}
\end{equation}
and
\begin{equation}\label{eq:shift-down-operator}
\hat{S}_{\text{bw}}:\begin{cases}
\begin{array}{rcl}\left|\uparrow,l\right> &\to& e^{i\varphi_{\uparrow}}\left|\uparrow,l-1\right>\\
\left|\downarrow,l\right>
&\to&e^{i\varphi_{\downarrow}}\left|\downarrow,l+1\right>\,,\end{array}\end{cases}
\end{equation}
where we use a short-hand notation of spin-position product states
$\left|s,l\right>\equiv\left|s\right>\otimes\left|l\right>$ with
$s=\{\uparrow,\downarrow\}$ and $l=0,\pm 1,\pm 2,\ldots$. The
subscript of $\hat{S}$ indicates the rotation direction of the
polarization, \textit{i.e.}, forward (fw) for $0\to\pi$ and backward (bw)
for $\pi\to 0$, shifting the spin-dependent potentials from one
overlapping configuration to another. A spin-dependent phase
$\varphi_{s}$ is accumulated during the shift with
$\varphi_{\uparrow}\not=\varphi_{\downarrow}$.

We implement the shift operators $\hat{S}_{\text{fw}}$ and
$\hat{S}_{\text{bw}}$ by feeding cosinusoidal driving ramps
$V_{\text{fw}}(t)\equiv V_{0}+(V_{\pi}-V_{0})[1-\cos(\pi\cdot
t/\tau)]$ and $V_{\text{bw}}(t)\equiv V_{\text{fw}}(\tau-t)$ to
the amplifier's input, where $\tau$ denotes the ramp time, and
$V_{0}$ and $V_{\pi}$ are the input voltages corresponding to the
$\theta=0$ and $\theta=\pi$ configuration, respectively. For ramp
times $\tau>14\,\mu\text{s}$, these ramps almost perfectly
translate to the rotation angle $\theta(t)$. The ramp time is
chosen such that excitations between vibrational states of the
atoms are negligible while still being sufficiently fast to finish
the experimental sequence within the phase coherence time of
$T_{2}\approx 1\,\text{ms}$ --- a mandatory requirement for advanced
applications of spin-dependent transport\cite{Man03a, Kar09}. We
calculate the excitation probability by using first-order
perturbation theory\cite{Hae01}, where the trapping potentials
are approximated as harmonic, and the axial and radial dynamics of
the atoms are assumed to be decoupled. For the cosinusoidal
driving ramps, we find an optimum ramp time of
$\tau=30\,\mu\text{s}$, resulting in an excitation probability of
less than $3\%$ per shift.

Spin-dependent transport over distances of $L=1,2,3,\ldots$
lattice sites to the left or to the right (\textit{i.e.},\ by $\pm
L\cdot\lambda/2$) is realized by alternating the applications of
the shift operators $\hat{S}_{\text{fw}}$ and
$\hat{S}_{\text{bw}}$ with $\pi$-pulses flipping the spin states
between ($\hat{U}_{\pi}:\left|s,l\right>\to i\left|-s, l\right>$),
see Fig.~\ref{fig2}:
\begin{equation}\label{eq:transport-sequence-operator}
\begin{array}{l}\hat{T}_{2L}\equiv
{\left(\hat{U}_{\pi}\hat{S}_{\text{bw}}\hat{U}_{\pi}\hat{S}_{\text{fw}}\right)}^{L} \vspace{5pt}\\
\hat{T}_{2L}\,:\begin{cases}\begin{array}{rcl}\left|\uparrow,l\right> &\to &(-1)^{L}e^{i(\varphi_{\uparrow}+\varphi_{\downarrow})L}\left|\uparrow,l+2L\right>\\
\left|\downarrow,l\right> &\to
&(-1)^{L}e^{i(\varphi_{\uparrow}+\varphi_{\downarrow})L}
\left|\downarrow,l-2L\right>\,.\end{array}\end{cases}
\end{array}
\end{equation}
The operator $\hat{T}_{2L}$ is intentionally constructed from an
even number of transport steps. By this, the entire transport
sequence always starts and ends in the $\theta=0$ configuration,
which, due to technical issues, is the experimentally most robust
configuration providing long-term stability on the time scale of
seconds. Such long-term stability is required for detection of the
initial and the final positions of the atoms in the lattice by
using fluorescence imaging\cite{Kar09a} with a typical exposure
time of $1\,\text{s}$. Otherwise, because the trap depth for
atoms, except those in $\left|F=4,m_{\text{F}}=4\right>$, is
reduced in the non-overlapping case, atom losses would be enhanced
during irradiation with the near-resonant light required for the
imaging process.

The $\pi$-pulses are driven by resonant rectangular microwave
pulses at $2\pi\times 9.2\,\text{GHz}$ with a pulse duration of
$8\,\mu\text{s}$ (Rabi frequency of $\approx 2\pi\times
60\,\text{kHz}$). Alternatively, broadband composite pulses, so
called $90_{0}225_{180}315_{0}$-pulses\cite{Sta85, Lev96} with a
duration of $24\,\mu\text{s}$ are employed to compensate for
possible errors in the pulse frequency or power. To minimize
frequency broadening from vectorial and tensorial contributions to
the differential light shift, microwave transitions between the
internalstates of the atoms are only induced in perfectly
overlapping trapping potentials. The microwave pulses are applied
after each shift operation with a time delay of $2\,\mu\text{s}$,
which allows lattice polarization transients to settle. The latter
are caused by the limited bandwidth of the polarization control
setup and by excitation of mechanical resonances of the EOM
crystal. The time delay has been determined using microwave
spectroscopy, utilizing the fact that spectra are significantly
broadened by any displacement of the $U_{\uparrow}$,
$U_{\downarrow}$ potentials. This method ensures a maximum
deviation of $|\Delta\theta|<10^{-2}$ for the rotation angle in
the overlap configuration, corresponding to a maximum relative
displacement of the spin-dependent trapping potentials of less
than $1\,\text{nm}$ from perfect overlap.

\section{Measurements and discussion}
To investigate the efficiency of our spin-dependent transport over
distances of several lattice sites, we load an average of eight
atoms into the optical lattice, which are randomly distributed
over a region of about 150 lattice sites. We determine the initial
positions of the atoms by fluorescence imaging (exposure time of
$1\,\text{s}$) and prepare them in the spin state
$\left|\uparrow\right>$ by using optical pumping. Details on
state-preparation and detection can be found in
Refs.~\citealp{Kar09a} and \citealp{Kar10}. We subsequently apply the
transport sequence defined by $\hat{T}_{2L}$ for
$L=\{1,2,\ldots,11\}$, each composed of an even number of $2L$
transport steps (see Eq.~\eqref{eq:transport-sequence-operator}).
After transporting the atoms, we determine their final positions
by fluorescence imaging and calculate their transport distances,
\textit{i.e.}\ the final positions relative to the initial ones, by
considering only those atoms that are initially sufficiently far
separated so that even in case of transport errors, their
transport paths cannot cross.
\begin{figure}
    \centering
    \includegraphics[width=0.8\columnwidth]{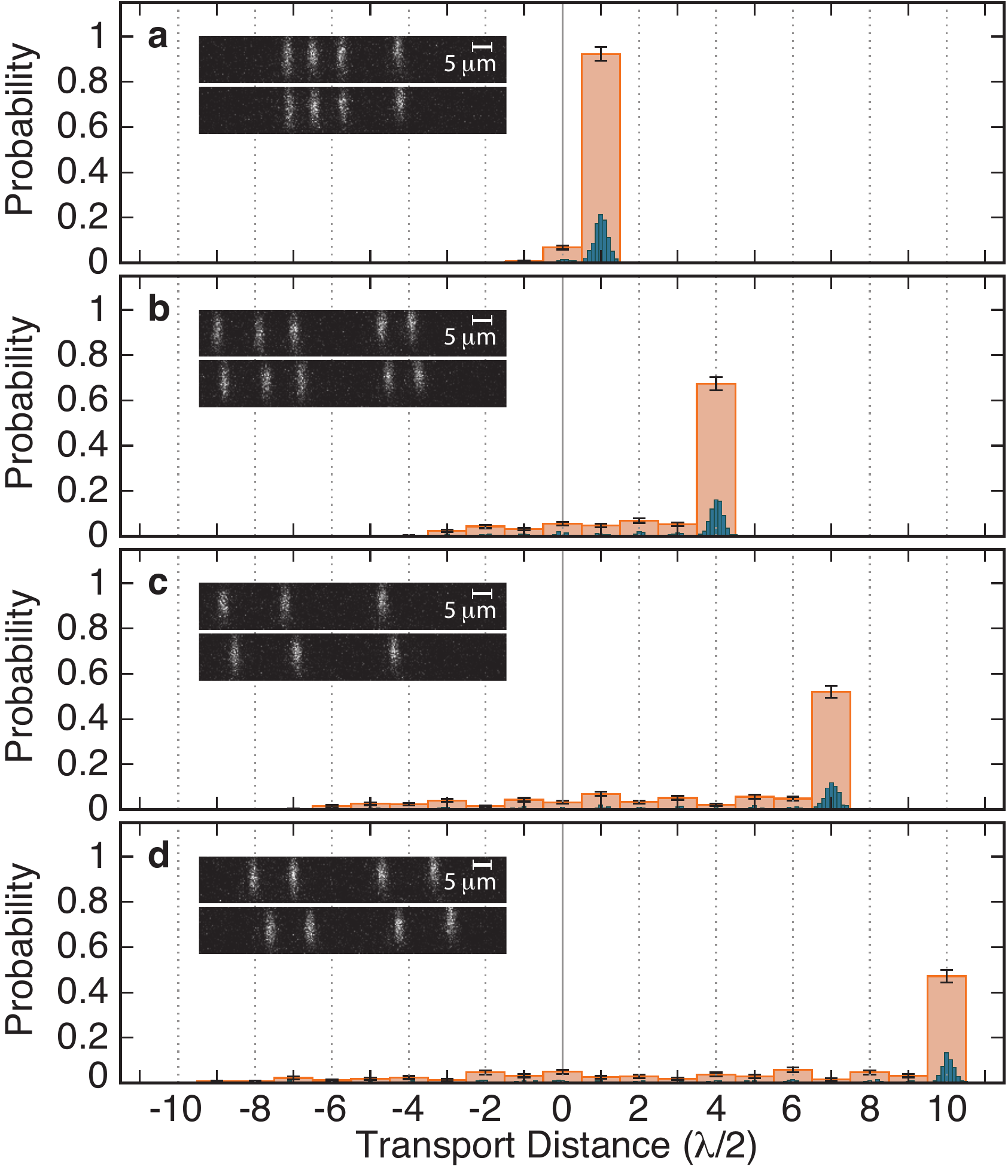}
    \caption{(Color online) Probability histograms of transport distances of atoms
    initialized in the spin state $\left|\uparrow\right>$ after a
    spin-dependent transport comprised $2L$ transport steps for
    (\textbf{a}) $L=1$, (\textbf{b}) $L=4$, (\textbf{c}) $L=7$ and
    (\textbf{d}) $L=10$. Digitized histograms (light shade) with a
    bin width of $\lambda/2$ indicate the transport distances in units
    of lattice sites. Densely sampled histograms (dark shade) with a
    bin width of $\lambda/20$ reveal the periodicity of the optical
    lattice, indicating the absence of significant lattice drifts
    during the transport sequence. Insets show initial (upper) and
    final (lower) fluorescence sample images of transported
    atoms.}\label{fig3}
\end{figure}
In Fig.  \ref{fig3}, probability histograms of transport distances
for different numbers of transport steps $2L$ are shown. Each
histogram has been normalized to the total number of analyzed
atoms ($500$ -- $1000$). Digitized histograms with a bin width of
$\lambda/2$ reveal the probability $P_{2j}$ of finding an atom
transported over a distance of $j\cdot\lambda/2$ ($j=0,\pm 1,\pm
2,\ldots$). Densely sampled histograms with a bin width of
$\lambda/20$ reveal Gaussian peaks centered at integer multiples
of $\lambda/2$. These peaks clearly reproduce the periodicity of
the optical lattice verifying the position resolution achieved in
our fluorescence detection and indicating that no significant
drifts of the optical lattice, relative to the imaging optics
occurred during the transport sequence. From the Gaussian peaks,
we estimate the reliability of inferring the correct transport
distance $j\cdot\lambda/2$, yielding $>99.7\%$.

In the ideal case, starting from state $\left|\uparrow\right>$ and
applying $2L$ transport steps, we expect only a single histogram
bar at transport distance $L\cdot\lambda/2$ with probability
$P_{2L}=100\%$. We, indeed, observe the majority of atoms at the
expected transport distance, however, with a reduced probability.
The transport efficiency for $2L$ transport steps is the measured
probability for an atom to arrive at the nominal transport
distance. Note that this definition does not contain any statement
regarding the coherence properties of the transport. It solely
reveals the successful displacement of atoms by $L$ lattice sites.

At a temperature of $10\,\mu\text{K}$ and a potential depth of
$k_{\text{B}}\times 80\,\mu\text{K}$, tunneling of atoms is
extremely improbable during shifts of the spin-dependent
potentials ($\approx 10^{-4}$ per shift, inferred from
band-structure calculation). For our analysis, we, therefore,
assume perfect shift operations $\hat{S}_{\text{fw}}$ and
$\hat{S}_{\text{bw}}$ and attribute the imperfections to the
preparation and the evolution of internal states of atoms.

\subsection{State Initialization and Photon Scattering}
Errors in the state initialization at the beginning or scattering
of photons from the light field of the optical lattice during the
transport sequence may transfer the atoms into spin states outside
the Hilbert space of the qubit, \textit{i.e.}, to
${\left|F=4,m_{\text{F}}\not=4\right>}$ or
${\left|F=3,m_{\text{F}}\not=3\right>}$. Except for
$m_{\text{F}}=0$, atoms in these states effectively still
experience a spin-dependent trapping potential due to an unequal
fraction of potential contributions $U_{+}(z,\theta)$ and
$U_{-}(z,\theta)$, however, with a reduced trap depth at
$\theta=\pi/2$. Hence, for atoms in states
$\left|m_{\text{F}}\right|<3$, we expect significant atom losses
during the transport sequence. More importantly, once the internal
states of the atoms have left the Hilbert space of the qubit,
their possible transition frequencies are far detuned from the
preset $\pi$-pulse frequency by at least $2\pi\times
1\,\text{MHz}$. These atoms are then no longer affected by the
microwave pulses of the transport sequence. For an even number of
transport steps, they are henceforth transported back and forth
close to the lattice site at which their internal state left the
Hilbert space of the qubit for the first time, staying behind the
nominal transport distance and thus reducing the transport
efficiency. In contrast to the scattering of photons, which can be
regarded as a per-step imperfection, state initialization is a one
time operation that is performed only once at the beginning of
each transport sequence. The corresponding error imposes,
therefore, merely an upper limit on the transport efficiency,
irrespective of the number of subsequent transport steps,
$P_{2L}\leq P_{\text{ini}}$.

In our experiment, state initialization is limited by the optical
pumping efficiency, yielding $P_{\text{ini}}>97\%$. The scattering
of photons from the light field of the optical lattice with a
calculated Raman and Rayleigh scattering rates of
$10\,\text{s}^{-1}$ and $5\,\text{s}^{-1}$, respectively, is
negligible at the time scale of a typical transport step with a
probability of the order of $10^{-4}$. The effect of photon
scattering on the transport efficiency is, thus, neglected.

\subsection{Perturbations of $\pi$-pulses}
Typical pulse errors can be divided into static and dynamic perturbations of pulse area, frequency and phase during the pulse. In ensemble averages, these perturbations usually contribute to inhomogeneous (static perturbations) or homogeneous (dynamic perturbations) broadenings or shifts. Their effect on the
transport efficiency $P_{2L}$ can be calculated by replacing the
ideal $\pi$-pulse operators $\hat{U}_{\pi}$ in
Eq.~\eqref{eq:transport-sequence-operator} by their perturbed
counterparts.

Static phase perturbations do not affect the transport efficiency,
because spin states of correctly transported atoms are
automatically prepared in the pure basis states
$\left|\uparrow\right>$ or $\left|\downarrow\right>$ by the shift
operations. The same holds for phases accumulated during the shift
so that the transport efficiency is insensitive to shift-induced
dephasing. Static perturbations of pulse area and frequency, on
the other hand, directly affect the spin-flip efficiency in each
step and, thus, the transport efficiency $P_{2L}$ as well. Such
perturbations, for instance, are induced by the radial oscillation
of atoms in the trapping potential, which in our case may be
regarded as frozen on the time scale of a single pulse\cite{Kuh05}. Similarly, the thermal distribution over the axial
vibrational states causes an inhomogeneous frequency broadening
and induces a static perturbation of the Rabi frequency and, thus,
of the pulse area via the Franck-Condon factor\cite{Foe09}.
Vibrational excitation during the shift operations may even
increase these inhomogeneities. From the transport efficiency
alone, however, static perturbations are not only
indistinguishable from one another but also from different
dynamical perturbations such as repeatable drifts or fluctuations
of lattice depth and polarization or magnetic fields. We,
therefore, subsume all these perturbations in an effective
spin-flip efficiency $\bar{P}_{\text{flip}}$ per pulse.

Assuming that $\bar{P}_{\text{flip}}$ does not significantly
change over the entire transport sequence, we expect the transport
efficiency to exponentially decay according to
\begin{equation}\label{eq:transport-error-model}
P_{2L}^{\text{err}}=P_{\text{ini}}\cdot{(\bar{P}_{\text{flip}})}^{2L-1}\,.
\end{equation}
The alignment procedure of the experimental setup aims at
identical microwave spectra for both overlapping configurations,
${\theta=0}$ and ${\theta=\pi}$, including the amplitudes,
positions and shapes of the resonance peaks, supporting this
assumption.

\begin{figure}
    \centering
    \includegraphics[width=0.8\columnwidth]{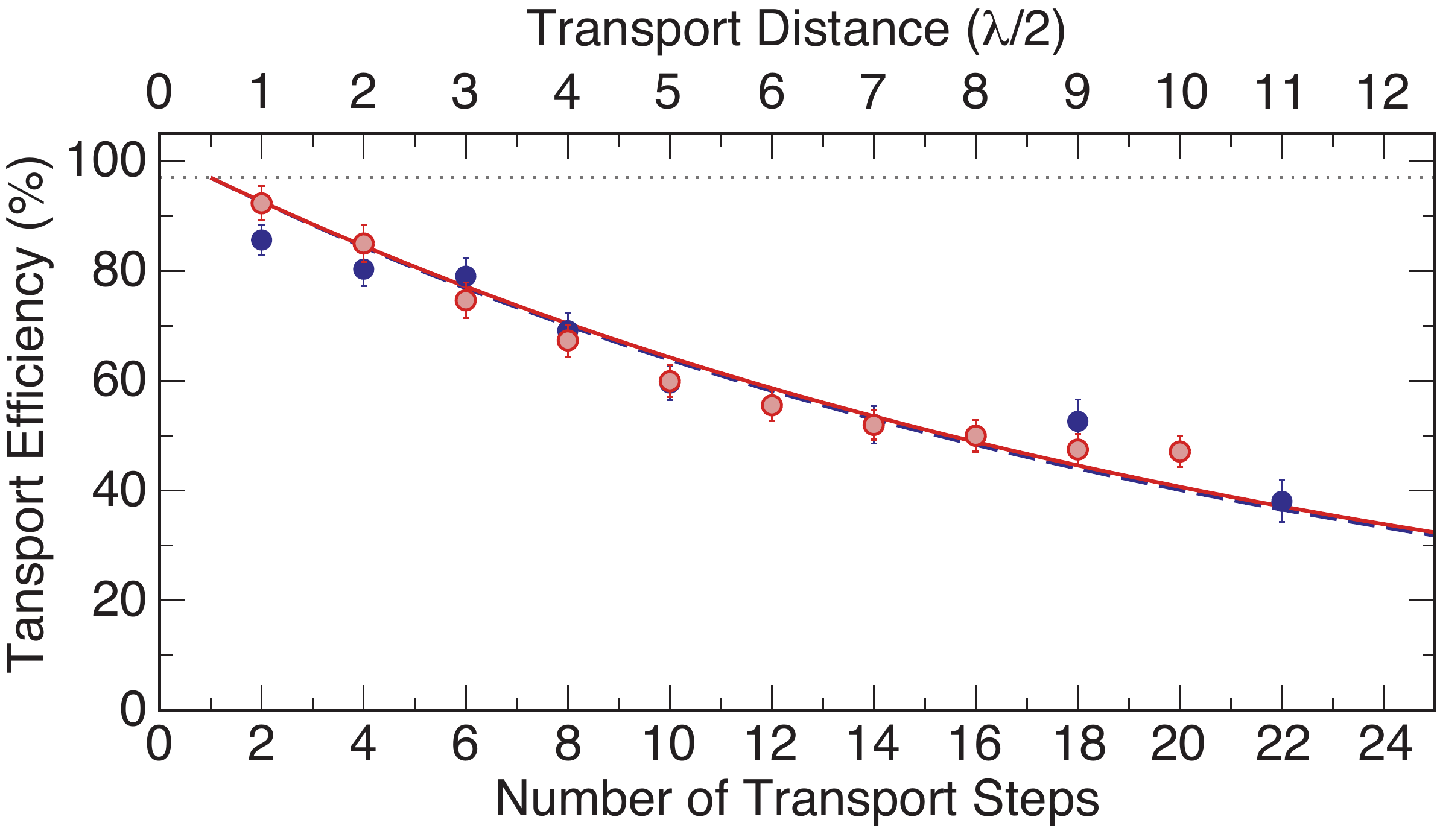}
    \caption{(Color online) Transport efficiency $P_{2L}$ as a function of the number of
    transport steps $2L$ for a sequence employing rectangular
    $180_{0}$($\pi$)-pulses ($\circ$) and composite
    $90_{0}225_{180}315_{0}$-pulses of equal Rabi frequency
    ($\bullet$). The vertical, dotted line indicates the upper
    transport efficiency limit imposed by the state initialization
    $P_{\text{ini}}$.  The solid (dashed) line shows a fit of the
    model function of Eq.~\eqref{eq:transport-error-model} to the data
    obtained for $180_{0}$-pulses ($90_{0}225_{180}315_{0}$-pulses) in
    the transport sequence.}\label{fig4}
\end{figure}
In Fig.~\ref{fig4}, the measured transport efficiency as a
function of the number of transport steps is shown. It fits well
the exponential dependence of
Eq.~\eqref{eq:transport-error-model}, yielding
\begin{equation}\label{eq:mean-trans-eff-rect}
    \bar{P}_{\text{flip}}=(95.5\pm 0.3)\%\,.
\end{equation}
Slight deviations of individual data points from the fit in
Fig.~\ref{fig4} could be partially attributed to slow drifts of
lattice polarization over time. Such drifts have been monitored
independently by measuring the polarization state of the
retro-reflected beam after passing the trapping region of the
atoms.

To reduce possible effects of static frequency and pulse area
perturbations on the spin-flip efficiency, including the
corresponding inhomogeneous broadening mechanisms, we have
replaced the regular, rectangular $\pi$-pulses by broadband
composite pulses in the transport sequence, namely
$90_{0}225_{180}315_{0}$-pulses. The latter are robust against
static detunings within a calculated
($P_{\text{flip}}\geq95\%$-)range of $|\delta|<2\pi\times
54\,\text{kHz}$ (compared to $|\delta|<2\pi\times 14\,\text{kHz}$
of the regular, rectangular $\pi$-pulses) for static pulse area
perturbations of up to $\pm 10\%$. The measured transport
efficiency as a function of the number of transport steps for this
case is shown in Fig.~\ref{fig4} ($\bullet$). From a fit of
Eq.~\eqref{eq:transport-error-model}, we obtain
\begin{equation}\label{eq:mean-trans-eff-cp}
    \bar{P}_{\text{flip},\text{CP}}=(95.5\pm 0.3)\%\,,
\end{equation}
which incidentally agrees with the value for the regular
$\pi$-pulses (see Eq.~\eqref{eq:mean-trans-eff-rect}). This means
that the compensation of static perturbations (\textit{e.g.},
inhomogeneities) by the composite pulses does not improve the
transport efficiency. This finding suggests that the finite
transport efficiency is dominated by dynamic perturbations, such
as lattice polarization drifts and fluctuations (decoherence)
during the pulse affecting the internal state of the atom. Also,
we cannot exclude that the improvement of the spin-flip efficiency
of the composite pulse is counteracted by additional dephasing
during the pulse due to its three times longer duration.

\subsection{Dynamic Perturbations and Decoherence}
Decoherence is typically characterized by a longitudinal
relaxation time $T_{1}$ and a dephasing time $T_{2}$.  We have
measured longitudinal relaxation by preparing the atoms in state
$\left|\uparrow\right>$ or $\left|\downarrow\right>$ and
performing a typical transport whereby $\pi$-pulses are removed
from the sequence, and thus their possible errors as well. Atoms
are then only transported back and forth to the initial lattice
site. Finally, we determine the population in state
$\left|\downarrow\right>$ after different times by using
state-selective push out\cite{Kar10}, yielding $T_{1}\approx
100\,\text{ms}$, in agreement with the calculated limit imposed by
the Raman scattering processes in the lattice. Measuring dephasing
during the pulses in the transport sequence and precisely
inferring the irreversible $T'_{2}$ time, \textit{e.g.}, from decay of Rabi
oscillations, turns out to be non-trivial due to technical
instability of the $\theta=\pi$ overlapping configuration on the
time scale required. We, therefore, estimate this $T'_{2}$ time by
solving the (pulse-driven) optical Bloch equations for the
measured spin-flip efficiency of
Eqs.~\eqref{eq:mean-trans-eff-rect} and
\eqref{eq:mean-trans-eff-cp}, yielding $T'_{2}\approx
100\,\mu\text{s}$. The corresponding spectrum, calculated from the
optical Bloch equations, deviates, however, from the measured
spectra in shape. These deviations may arise from repeated
(non-random) dynamic perturbations that are not covered by the
simple relaxation model in the Bloch equations. Consequently,
these perturbations should not be interpreted as fluctuations, and
the true irreversible dephasing time $T'_{2}$ may be much longer
then estimated above. In comparison, in a ``static'' lattice
configuration (no transport), this dephasing time $T'_{2}$ is on
the order of several milliseconds, as inferred from Rabi
oscillations. We conclude from this fact that the spin-flip
efficiency in the transport sequence is, indeed, reduced by
dynamic perturbations, probably caused by drifts or oscillations
of the lattice polarization, which possibly have not yet fully
decayed after each shift, when the pulses are applied. On the one
hand, such dynamic perturbation directly translate into drifts and
fluctuations of differential light shifts, leading to homogeneous
broadening of the transition frequency and, thus, dephasing. On
the other hand, they translate into drifts and fluctuations of the
Rabi frequency and, thus, of the pulse area via the Franck-Condon
factor\cite{Foe09}. Our interpretation is, furthermore, supported
by the higher spin-flip efficiency of $(98.9\pm 0.2)\%$ measured
in a ``static'' optical lattice. Employing optimum control
techniques\cite{DeC08}, either by microwave pulse-shaping or
active polarization control, might help to counteract or
compensate for dynamical drifts of our system,  and, thus, improve
the pulse efficiency in the future.

\section{Summary}
\begin{table}\centering
\begin{tabular}{llr}
    \toprule
    & Error source & Probability\\
    \colrule
    \multicolumn{2}{l}{A. One time errors}\\
    \colrule
    & Transport distance detection    & $<0.3\%$\\
    & State initialization            & $<3\%$\\
    \colrule
    B. &  Per step errors\\
    \colrule
    & Tunneling of atoms      & $0.01\%$\\
    & Raman scattering        & $0.01\%$\\
    & Imperfection of $\pi$ pulses    &\\
    & -- Regular rectangular pulses   & $<4.5\%$\\
    & -- Broadband composite pulses   & $<4.5\%$\\
    & -- For comparison: no transport    & $<1.1\%$\\
    & Vibrational excitations & $<3\%$\\
    \botrule
    \end{tabular}\caption{Compilation of error
probabilities (shift + pulse) limiting the measured transport
efficiency or affecting the quantum state
fidelity.}\label{tab:infidelities}
\end{table}
We have directly observed and analyzed the transport efficiency of
a unidirectional spin-dependent transport of single cesium atoms
in a 1D optical lattice. Relevant error sources are compiled in
Table~\ref{tab:infidelities}. For our experimental parameters,
tunneling of atoms is negligible during the transport, and
excitations between vibrational states are minimized by
numerically finding an optimum ramp time for a chosen ramp. Photon
scattering processes and errors in state initialization play
marginal roles while successful displacement of the atoms is
insensitive to dephasing during the shift. The transport
efficiency is mainly limited by the evolution of the internal
states of the atoms. Our analysis of static and dynamic pulse
perturbations suggests that spin-flip efficiencies are limited by
repeated drifts or fluctuations during the transport operations.

\begin{acknowledgments}
The authors gratefully acknowledge financial support by the
Deutscher Akademischer Austauschdienst exchange program (KOSEF),
the Deutsche Forschungsgemeinschaft FOR635, Me971/25-2, and the
European Commission (IP AQUTE). M.~Karski acknowledges support
from the Studienstiftung des deutschen Volkes, J.~M.~Choi received
partial support from a Korea Research Foundation grant funded by
the Korean Government (Ministry of Education, MOEHRD).
\end{acknowledgments}

\end{document}